# High-temperature ferromagnetism and ferroelasticity in ultraflexible atomically thin square-shaped lattices

Xinyuan Huang,[1,#] Yueqiao Qu,[2,#] Yu Liao,[1] Qian Zheng,[1] Ran Liu,[3] Yu Chen,[4] Liang Liu,[5] Junzhong Wang,[1] and Gang Yao[1,*]

[1]School of Physical Science and Technology, Southwest University, Chongqing 400715, China
[2]School of Physics and Astronomy, Shanghai Jiao Tong University, Shanghai 200240, China
[3]Department of Physics, Beijing Normal University, Beijing 100875, China
[4]School of Science, Inner Mongolia University of Technology, Hohhot 010051, China
[5]School of Physics, State Key Laboratory for Crystal Materials, Shandong University, Jinan 250100, China

[#]These authors contributed equally: Xinyuan Huang, Yueqiao Qu
*Corresponding author. Email: yaogang@swu.edu.cn

The coexistence of high-temperature intrinsic ferromagnetic ordering, large magnetic anisotropy, along with novel mechanical properties such as ferroelasticity and flexibility, in experimental feasible two-dimensional (2D) crystals is greatly appealing for nanoscale spintronics. However, the progress in identifying such materials is limited. Here, by first-principles calculations, we report the findings of an extraordinary combination of the above qualities for the first time in a new 2D exfoliated FeSi nanosheet in the *P*4/*nmm* space group. Due to the strong anion-mediated superexchange interaction, the monolayer FeSi (ML-FeSi) exhibits a Curie temperature $T_c$ as high as 830 K, surpassing the current experimental record (344 K for ML-$Cr_3Te_4$). Furthermore, including FeSi, such isostructural lattices all demonstrate exceptional softness, as evidenced by their ultra-low in-plane stiffness. Remarkably, the transition metal atom and square-shaped crystal form work together to give this family of ML materials unique properties that can transition from Ising-like 2D ferromagnets in FeSi, MnP, MnAs, CrP, FeI, and VAs to 2D-XY ones in CrAs, VP, and multiferroic MnGe and TiTe. Overall, our work highlights such 2D lattices as promising candidates in emerging multifunctional device applications and nontrivial topological spintronics.

  The discovery of spintronic phenomena, such as the giant magnetoresistance effect, has opened new avenues for manipulating the spin and charge of materials [1-3]. However, the spins of actual materials possess a three-dimensional (3D) degree of freedom. Breaking the continuous rotational spin symmetry can lead to the emergence of various spontaneous magnetic orders by defying the Mermin-Wagner theorem [4,5].

For example, in 2D structures, the long-range FM order can persist at finite temperatures if a particular easy axis is introduced [6-9]. For practical spintronic devices capable of operating at room temperature, atomically thin 2D ferromagnets with a combination of large magnetic anisotropy energy (MAE) and strong FM coupling are fundamental. Conversely, a 2D material that has an easy magnetization plane, also known as a 2D-XY ferromagnet [10,11], leads to a residual SO(2) symmetry. This extra degree of freedom offers hope for a range of intricate topological spin states, such as merons, skyrmions, and magnetic bubbles, which have sparked significant research enthusiasm [12-16]. The first experimental 2D-XY ferromagnet achieved is $CrCl_3$ grown on a graphene/6H-SiC(0001) substrate [17]. However, its Berezinskii-Kosterlitz-Thouless (BKT) transition temperature $T_{BKT}$ is rather low (~12.95 K), severely limiting its further research in the laboratory. To this end, having a combination of large magnetic anisotropy energy and strong FM coupling in 2D materials would be essential for advancing both fundamental physics and technological applications.

Beyond conventional 2D ferromagnetism, multiferroic materials—exhibiting multiple ferroic orders such as antiferromagnetic (AFM) or FM, ferroelastic (FEL), and ferroelectric (FE)—represent a transformative paradigm in spintronics and electronics [18]. These materials offer the potential to replace traditional silicon-based technologies, driven by the demand for highly compact multifunctional information devices and their promising applications in nonvolatile memories with high integration densities. Because FE and FM orders have conflicting origins, there is a scarcity of magnetoelectric materials, prompting a strong focus on AFM/FM-FEL multiferroicity. The most notable examples are hole-doped α-SnO and α-PbO MLs [19,20], opening the door for FM-FEL multiferroicity at atomic thickness in a single material at the 2D limit. However, in addition to the low $T_c$, their FM feature is limited to a specific range of hole carrier densities. Theoretically, 2D tetragonal magnets have demonstrated the advantages of rectangular lattice for novel multiferroic applications due to their crystal anisotropy [21-26], but quite a few of them have high FM $T_c$ and FEL simultaneously, greatly restricting their practical applications. Nevertheless, despite tremendous efforts, seeking FM-FEL materials with strong FM coupling is still essential and challenging. Also, 2D FEL materials inherently possess low in-plane stiffness, which is a critical aspect for flexible functional devices, but has received limited attention in the past.

The 2D square-shaped lattice in the space group *P4/nmm* (No. 129) has recently garnered significant attention as a rich topological and magnetic properties have been revealed in this material family, typically ML-FeSe and its derivatives. These include, for example, interface superconductivity [27], connate topological superconductivity [28,29], high-temperature quantum anomalous Hall (QAH) effect [30-32], and AFM-FEL multiferroicity [33]. Remarkably, thanks to the experimental synthesis of layered 3D bulk $ATM_2X_2$ crystals (e.g., A=Ba, Ca, Th, Zr, etc; TM=Fe, Cr, Mn, V; X=Ge, Si, P, As) that crystallize with the space group *I4/mmm* and have weak van der Waals (vdWs) interlayer interactions, featuring an alternating stacking structure of TMX and alkaline earth A layers along the z-axis (Fig. S1 in the Supplemental Material [34]). Possible intrinsic AFM/FM in their TMX exfoliated nanosheets was theoretically suggested [35-38]. Motivated by research progress, we performed systematic first-principles calculations for such a type of TMX lattice, in light of their existing bulk phases with varying transition-metal elements. Remarkably enough, in this work, we report such a 2D lattice in FeSi, a brand-new phase of 2D FeSi$_x$ alloys [39,40], which surprisingly exhibiting room-temperature long-range FM order, robust FM-FEL multiferroicity, and exceptional flexibility. We show that ML-FeSi has great dynamical and thermal stability and can be fabricated, for example, by exfoliating existing 3D bulk with a relatively low cleavage energy comparable to that of MXenes. More interestingly, through a comprehensive examination of the reported isostructural lattices, we demonstrate that these materials are all unexpectedly flexible, with some members possessing isotropic easy-plane magnetization and/or FM-FEL multiferroicity.

Our first-principles calculation were performed using the projector augmented wave method [41,42] combined with density functional theory (DFT) in the Vienna ab initio simulation package (VASP) [43]. The wave functions in plane waves are expanded using a plane wave basis set with a cutoff energy of 500 eV. We adopted the generalized gradient approximation (GGA) in the Perdew-Burke-Ernzerhof (PBE) scheme to describe the exchange-correlation terms [44]. The first Brillouin zone (BZ) is sampled with 15×15×1 Monkhorst-Pack k-point meshes. A vacuum space of 20 Å is employed to avoid the interactions between periodic images along the *c* direction. Both the lattice constants and ionic positions are fully relaxed until the Hellman-Feynman forces on each atom are less than $10^{-2}$ eV/Å. The spin-orbit coupling (SOC) interaction is included in the calculation of MAE, while a *k*-mesh of 19×19×1 totaling 361 *k*-points is used in the BZ. The DFT+U method, developed by Dudarev et al., was used to treat the strong correlation effect related to the 3*d* orbitals of transition metal. The effective Hubbard $U_{eff}$ for Fe in monolayer FeSi is set to 3 eV, in line with previous studies [39,40]. For instance, a test for different $U_{eff}$ values within a reasonable range (0 to 4 eV) is also performed, which yields qualitatively similar results. Volumetric and charge density data were illustrated with the VESTA program [45]. The calculations of phonon dispersions were carried out with the finite displacement method by using the PHONOPY package [46]. To determine the thermal stability of the structure, *ab initio* molecular dynamics (AIMD) simulations were performed by using a 5×5×1 supercell for FeSi containing 100 atoms at 300 K under the canonical ensemble (NVT). The temperature is controlled using the Andersen thermostat. Monte Carlo simulations based on the classical Heisenberg spin Hamiltonian were performed with the Wolff algorithm to evaluate the FM $T_c$. In the simulation steps, a supercell consisting of a 20×20 square-shaped spin lattice with periodic boundary conditions is used, and the spins on all these sites are randomly flipped in all directions.

As shown in the inset of Fig. 1(a), this 2D material family has a square triple-layered lattice with two TM and two X atoms per primitive cell, in which the TM atoms are arranged in a square planar lattice, while the X atoms are puckered above and below the TM plane. This lattice is characterized by four main structural parameters: the in-plane lattice constant, the buckling height $\delta$ between the bilayer Si, the Fe-Si-Fe bond angle $\theta$, and the Fe-Si bond length *l*. Extracting a FeSi layer from an existing AFM bulk (Table SI), the optimized 2D lattice has smaller values in $a$ (3.60 Å), $\delta$ (1.52 Å), and $\theta$ (81.3°), but a larger value in $l$ (2.36 Å). Therefore, this highly buckled (HB) 2D structure may weaken the Fe-Si interaction, resulting in a possible strain response. As reported, the SrTiO$_3$ (STO) substrate has been extensively selected in the fabrication of Fe-based superconductor films through molecular beam epitaxy (MBE), while the as-grown sample is usually subject to a certain level of strain due to lattice mismatches [27,47]. Tensile strain-induced isostructural and magnetic phase transitions have also been theoretically predicted in ML-MoN$_2$ [48]. Therefore, it is necessary to consider the effect of in-plane biaxial strain (epitaxial strain) on the structure to find a stable phase for FeSi nanosheets. We calculate the binding energy of the optimized structure under a large external strain. As shown in Fig. 1(a), three isostructural lattice forms were found: the HB, low-buckled (LB), and planar (PL) phases, indicated by distinct local energy minima, with relative energies of 0, 31, and 33 meV/atom, respectively.

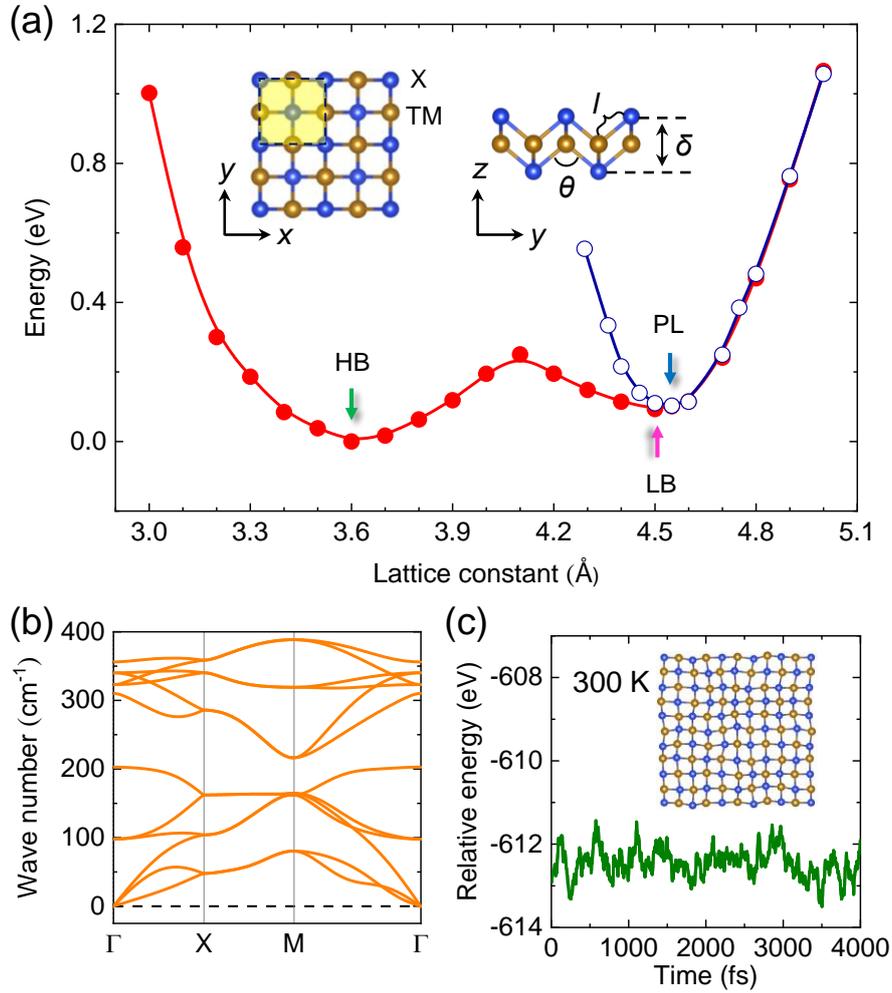

FIG. 1. (a) Relative energy versus lattice constants of three isostructural 2D FeSi crystals. The initial optimized high-buckled (HB), low-buckled (LB), and planar (PL) phases are indicated by green, pink, and blue arrows, respectively. The lattice constants corresponding to these phases are 3.60 Å, 4.51 Å, and 4.56 Å. The inset displays a ball-and-stick representation of the puckered geometry from top and side views, with the primitive cell delineated by the yellow-shaded region. (b) Phonon dispersion and (c) AIMD simulations of FeSi in the HB-phase.

The experimental fabrication of 2D crystals is heavily relies on their stability. The thermal and dynamic stabilities of the possible structures of FeSi MLs are assessed by phonon and AIMD trajectories. For the HB phase, no imaginary vibrational mode is found [Fig. 1(b)], implying its lattice-dynamical stability. However, the presence of imaginary frequencies in the remaining two phases indicates their dynamical instabilities (Fig. S2 in the Supplemental Material [34]). Therefore, the focus of the investigation will be on the 2D FeSi crystal in the HB-phase, which is referred to as ML-FeSi. Figure 1(c) displays the final snapshot of the ML-FeSi at 300 K for 4000 fs from AIMD simulations. The 2D planar network is well-maintained without broken bonds or phase transitions, suggesting that the structure of ML-FeSi is thermally stable at room temperature. This is reinforced by the small fluctuations observed in the simulated total energy-time curve. To further evaluate the structural stability of FeSi, we calculated its cohesive energy using the formula $E_{coh} = (2E_{Fe} + 2E_{Si} - E_{FeSi})/4$, where $E_{FeSi}$, $E_{Fe}$, and $E_{Si}$ denote the energy of FeSi unit cell, and isolated Fe and Si atoms, respectively. The calculated $E_f$ of ML-FeSi is 4.22 eV/atom, which is comparable to or larger than 4.18 eV/atom for ML-FeSe and 3.60~4.13 eV/atom for the previously studied 2D FeSi$_x$ alloys, bulk Fe$_2$Si, Fe$_3$Si, α-FeSi$_2$, and Fe-Si alloys [39,40], suggesting that the ML-FeSi is structurally stable.

Given the strong stability of ML-FeSi and the experimentally synthesized layered bulk AFe$_2$Si$_2$ crystals, it is expected to explore the possibility of obtaining a ML of FeSi from the bulk. In terms of thermodynamics, the exfoliation process needs to exceed a cleavage energy, $E_{cl}$, which is determined by the strength of the interlayer coupling [49]. The most common approaches used for exfoliation are mechanical cleavage and liquid exfoliation. Taking AFe$_2$Si$_2$ (A=La, Pr, Yb, and Eu) as an example, we found that as the interlayer distance, $d$, increases, the relative energy saturates quickly to a value corresponding to $E_{cl}$, ranging from 0.16 eV/Å$^2$ for EuFe$_2$Si$_2$ to 0.23 eV/Å$^2$ for LaFe$_2$Si$_2$ (Fig. S3 in the Supplemental Material [34]). These values are comparable to those of MXenes (0.086~0.205 eV/Å$^2$)[50] and 2D non-vdWs systems (0.31 eV/Å$^2$) [40,51], indicating that there is a high likelihood of obtaining an ML of FeSi by exfoliating its bulky samples. Especially, the theoretical cleavage strength σ obtained by the maximum derivative of $E_{cl}$ ($d$) is about 0.8 GPa, which is even lower than those of the typical 2D materials MnPSe$_3$ and graphene (1.2~2.1 GPa) [52,53], reinforcing again its superior experimental feasibility.

To determine the ground state, a 4×4×1 supercell of FeSi with different magnetic configurations shown in Fig. 2(a) is constructed. Interestingly, FeSi favors the FM ground state, with a net magnetic moment on Fe atoms of about 2.0 μ$_B$. The FM state is 413, 331.9, and 571.6 meV/Fe lower in energy compared to the three AFM states (AFM$_1$, AFM$_2$, AFM$_3$), respectively. The energy difference between FM and AFM$_1$ is higher than that of the most studied CrX$_3$ (X=Cl, Br, I) MLs [16,54,55], implying a strong FM exchange coupling.

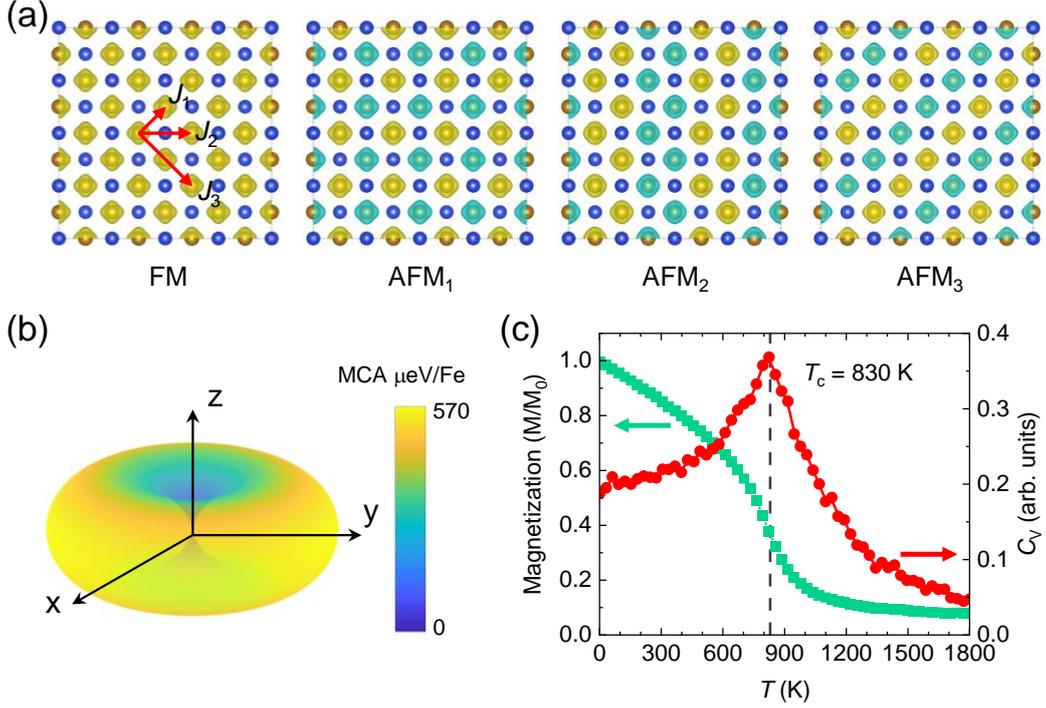

FIG. 2. (a) Four possible magnetic configurations considered for ML-FeSi. The spatial spin density for the Fe atoms is also shown, with up spin and down spin colored in yellow and blue, respectively. The isosurface value is taken at 0.05 e/Å$^3$. The solid red arrows indicate the exchange paths of the exchange parameters. (b) Angular dependence of the MCA energy of ML-FeSi with the direction of magnetization lying in the entire space. (c) Specific heat ($C_v$) and normalized magnetization as functions of temperature from MC simulations.

Another key issue for spintronics is the MAE, which quantifies how energy varies with magnetization orientation and is crucial for the thermal stability of magnetic data storage, while a high MAE implies the potential to reduce the grain size for each data bit. Due to the relativistic effect, MAE arises primarily from two origins: magnetocrystalline anisotropy (MCA), driven by SOC, and magnetic shape anisotropy (MSA), related to the Breit modification of the relativistic two-electron energy. To evaluate the MCA, we assess the influence of the spin direction of the Fe atom on energy levels when considering SOC. Figure 2(b) illustrates the angular dependence of the FM MCA energy for ML-FeSi. The strongly out-of-plane angle-dependent MCA and an out-of-plane easy axis can be clearly observed. The MCA energy, defined as the energy difference (with SOC included) between the in-plane and out-of-plane polarizations (i.e., MCA=$E_{\parallel}^{SOC} - E_{\perp}^{SOC}$), is calculated to be 570 μeV/Fe, implying a large SOC. On the other hand, MSA (MSA=$E_{\parallel}^{D-D} - E_{\perp}^{D-D}$) can be achieved through classical magnetic dipole-dipole (D-D) interactions, where the energy of such interaction is expressed as [56]

$$E_{D-D} = -\frac{1}{2}\frac{\mu_0}{4\pi}\sum_{i \neq j}\frac{1}{\mathbf{r}_{ij}^3}\left[\mathbf{M}_i \cdot \mathbf{M}_j - 3\frac{(\mathbf{r}_{ij}\cdot\mathbf{M}_i)(\mathbf{r}_{ij}\cdot\mathbf{M}_j)}{\mathbf{r}_{ij}^2}\right]. \quad (1)$$

Here, $\mu_0$ denotes the vacuum permeability, while $\mathbf{M}_i$ and $\mathbf{M}_j$ are the local magnetic moments separated by a distance $\mathbf{r}_{ij}$. Such an interaction typically promotes magnetization along the in-plane direction. A small but positive MSA value of 11 μeV/Fe is obtained for FeSi. This yields an overall MAE of 580 μeV/Fe, indicating the out-of-plane long-range FM ordered low-temperature phase, known as the Ising-like FM order, in ML-FeSi. The MAE of ML-FeSi is much larger than 1.4~65 μeV/atom for cubic Ni, Co, and Fe [57,58], providing it with significant advantages in magnetoelectronic applications.

For practical spintronics applications, a comprehensive understanding of the magnetic behavior at finite temperatures and the presence of room-temperature ferromagnetism is essential. The magnetic interactions in this system are described using the Heisenberg model along with a magnetic anisotropic term. The spin Hamiltonian, incorporating exchange interactions up to the third nearest neighbors, can be expressed as

$$H = -\sum_{i,j} J_1 \mathbf{S}_i \mathbf{S}_j - \sum_{i,j} J_2 \mathbf{S}_i \mathbf{S}_j - \sum_{i,j} J_3 \mathbf{S}_i \mathbf{S}_j - A(S_i^z)^2, \quad (2)$$

where $|\mathbf{S}|=1$, $J_1$, $J_2$, and $J_3$ represent the isotropic Heisenberg exchange interactions between nearest-neighbor (N), second (NN), and third (NNN) spins. A positive (negative) value of $J$ refers to the FM (AFM) coupling. $S_i^z$ is the spin component along the z (out-of-plane) direction, and $A$ describes the single-site magnetic anisotropy obtained by MAE. The exchange interaction parameters are determined to be $J_1$=49.84 meV, $J_2$=45.01 meV, and $J_3$=-6.42 meV, as detailed in Section IV of the Supplemental Material [34]). The significantly larger positive values of $J_1$ and $J_2$ compared to $J_3$ suggest their primary role in the FM arrangement of ML-FeSi. Furthermore, $J_1$ and $J_2$ are more than three times greater than that of CrI$_3$ (9 meV), hinting at a potentially high $T_c$. Through standard Wolf MC simulations, we estimate the $T_c$ of FeSi as 830 K [Fig. 2(c)], significantly exceeding the highest known $T_c$ (334 K) of the experimentally available ML-Cr$_3$Te$_4$ [8]. (See Fig. S4 in the Supplemental Material [34] for results considering different exchange interactions.) Note that the high FM $T_c$ of ML-FeSi is independent of the chosen computational approach (Table SII in the Supplemental Material [34] ). Taking ML-CrI$_3$ as a benchmark, a $T_c$ of 50 K is obtained by the same method (Fig. S5 in the Supplemental Material [34]), which agrees well with the measurements ($T_c$~45 K) [6], and predictions (42~70 K) [16,54,55].

The major mechanism and origin of magnetism in ML-FeSi can be elucidated from its crystal and electronic structures. As illustrated in Fig. 3(a), the distances between Fe$_1$ and Fe$_2$ (3.13 Å) and between Fe$_2$ and Fe$_3$ (3.6 Å) in ML-FeSi are significantly larger than the 3$d$ electron orbital radius (~0.48 Å), resulting in weak Fe-Fe direct exchange. However, the bonding angles, specifically 65° for Fe$_1$-Si-Fe$_2$ and 99° for Fe$_1$-As-Fe$_3$, both approach 90°, giving rise to a strong FM superexchange interaction between Fe atoms according to the Goodenough-Kanamori-Anderson (GKA) rules [59-61]. Thus, the competition between the Fe-As-Fe superexchange and Fe-Fe direct exchange interactions results in a large net FM value in $J_1$ and $J_2$. Regarding $J_3$, due to the various Fe-Si-Si-Fe superexchange paths that can lead to either FM or AFM coupling, the overall $J_3$ is weak and predominantly AFM. Figure 3(b) shows the band dispersion and density of states of ML-FeSi, revealing a metallic nature with a substantial bandwidth (exceeding 9 eV) and notable spin splitting in the Fe-3$d$ orbitals. Combining this feature with the plotted profiles of spin charge density depicted in Fig. 2(a), the magnetism of ML-FeSi is predominantly attributed to the spin-polarized Fe-3$d$ orbitals.

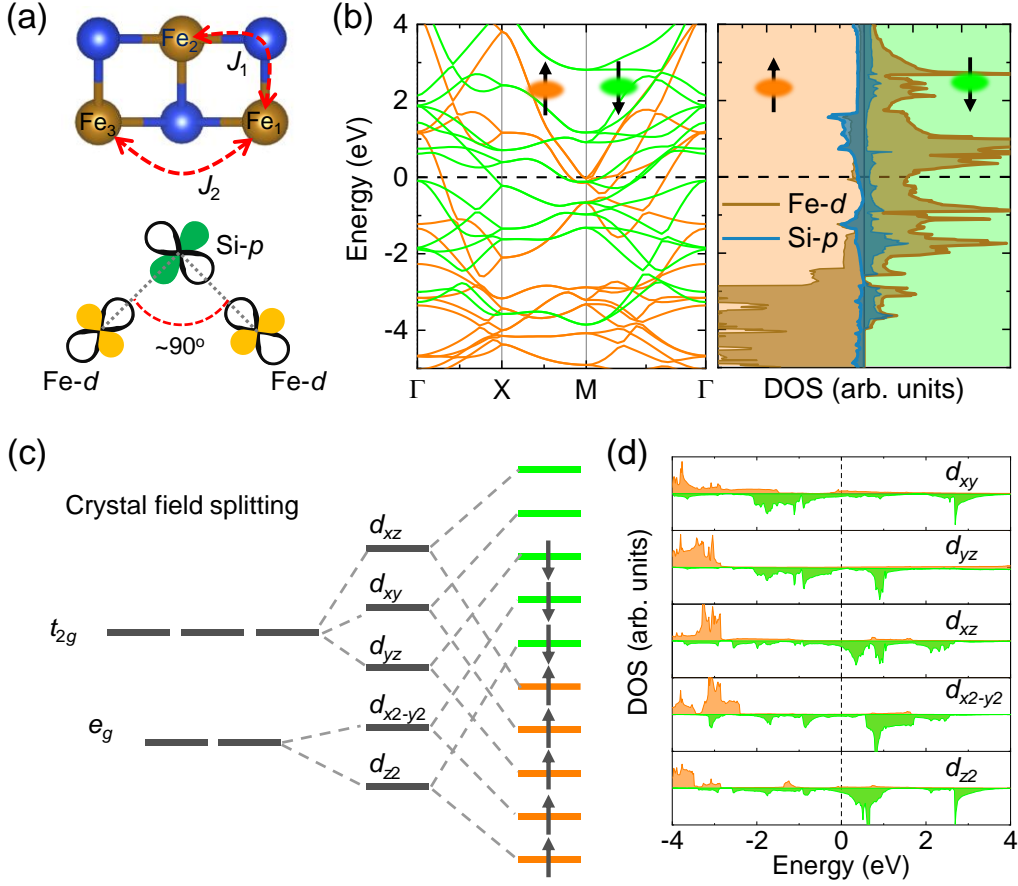

FIG. 3. (a) Schematic mechanisms of the Fe-Fe direct-exchange and Fe-As-Fe superexchange interactions. Here, $J_1$ and $J_2$ refer to the NN and NNN interaction parameters, respectively. (b) Band structure and projected density of states of ML-FeSi. The symbols "↑" and "↓" refer to up and down spins, respectively. (c) Crystal field splitting from an ideal tetrahedron to a distorted tetrahedron and the occupation of the Fe-3$d$ orbitals. (d) Orbital-projected DOS. The Fermi level is set at 0 eV.

Crystal field theory and charge transfer offer further insight into the local magnetic moment on Fe (2 $\mu_B$). The ideal tetrahedron crystal field induces a splitting of the energies of the five 3$d$ orbitals into two groups: the threefold $t_{2g}$ ($d_{xy}$, $d_{yz}$, and $d_{xz}$) and the twofold $e_g$ ($d_{x^2-y^2}$ and $d_{z^2}$) manifolds. In ML-FeSi, the presence of four Si atoms surrounding each Fe atom results in a distorted edge-sharing tetrahedral configuration, causing a reduction in symmetry and the consequent lifting of energy level degeneracy. As schemed in Fig. 3(c), the 3$d$ orbitals undergo further splitting into five non-degenerate orbitals, as supported by the orbital-resolved density of state (DOS) [Fig. 3(d)]. The valence electronic configuration of an isolated Fe atom is 3$d^6$s$^2$. Bader charge analysis suggests that Si atoms transfer 0.08 electrons to each Fe atom, which has almost no effect on the total number of electrons in the outer shell of an Fe atom. With a total of 8 electrons, these will fill the 10 nondegenerate orbitals sequentially, with the $d_{xz}$ and $d_{xy}$ ones half-occupied in a high-spin state, resulting in a net magnetic moment of 2 $\mu_B$ on Fe atoms. Also, the spin polarization near the Fermi level is mainly contributed by $d_{z^2}$ and $d_{xz}$ orbitals, agreeing with the above discussion.

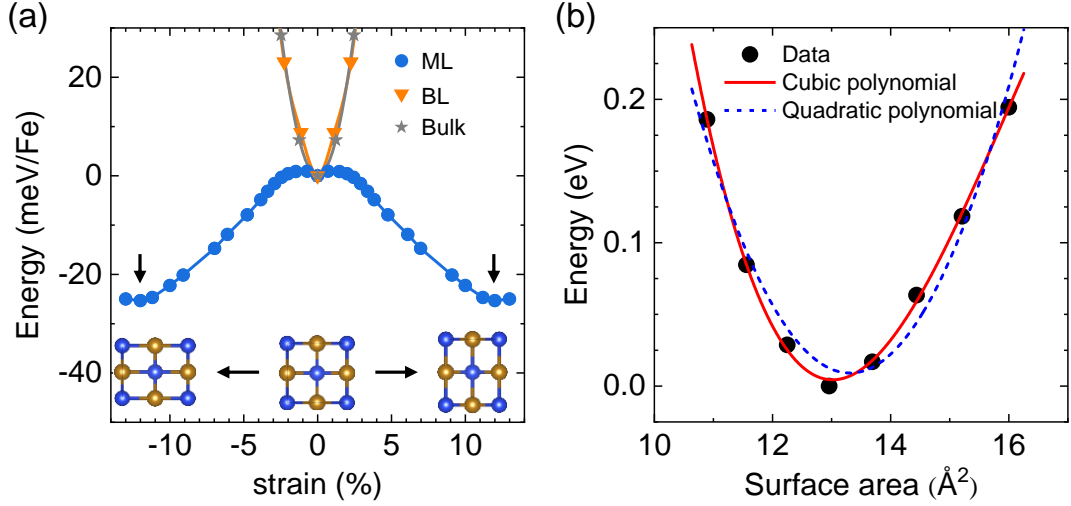

FIG. 4. (a) Elastic energy per atom of FeSi ML, bilayer, and bulk LaFe$_2$Si$_2$ as a function of in-plane antisymmetric diagonal strain. Two minima are indicated by arrows, and all energies are referenced to their ground state. The inset shows the FeSi unit cell with a schematic representation of the in-plane antisymmetric diagonal strains. (b) Variation of strain energy per unit cell with surface area for ML-FeSi under biaxial strain. The data are fitted with quadratic (blue dashed line) and cubic (red solid line) polynomials.

Although there is an overlap between the Fe $d$ orbitals and the Si $p$ orbitals at a large energy scale, the small difference in intensity between them implies weak $d↔p$ hybridization, potentially leading to FEL properties [62]. To validate this hypothesis, we performed elastic energy calculations on ML-FeSi with in-plane antisymmetric diagonal strain ($\varepsilon_{xx}$, $\varepsilon_{yy}$), involving compressive strain along one axis and tensile strain along the other, and relaxed the internal coordinates. Our results depicted in Fig. 4(a) reveal two degenerate minima at (+12, -12)% and (-12, +12)%, symmetrically positioned around the ground state, being characteristic of ferroelasticity. On the contrary, the bilayer FeSi and bulk ThFe$_2$Si$_2$ only have a minimum at $\varepsilon_{xx}$=-$\varepsilon_{yy}$=0. Thus, only the ML possesses FEL order with an intrinsic strain $\varepsilon_{xx} = -\varepsilon_{yy} = \pm 12\%$, establishing it as a multiferroic material with coexisting FM and in-plane FEL. Analogous results were reported in α-SnO, α-PbO, and MnGe MLs [19,20,38], which bear structural resemblances to ML-FeSi.

Previously, external strain has been demonstrated to influence the MAE and strength of magnetic coupling in various studies [22,48,54,63]. Here, we also investigated the response of strain on both the exchange energy ($E_{ex}= E_{FM}-E_{AFM}$) and the MAE of ML-FeSi. It is interesting that either a biaxial deformation within the range of -4% to 4% or an FEL strain does not disrupt the long-range FM order (Fig. S6 in the Supplemental Material [34]). Furthermore, given the prevalent use of STO substrate in the fabrication of FeSe or other non-vdW iron-based materials [27-29], the spin properties of ML-FeSi on STO (with a lattice mismatch of ~7%) are examined. The intrinsic long-range ferromagnetic ordering above room temperature is effectively preserved (Fig. S7 in the Supplemental Material [34]). These results strongly advocate the distinct advantages of ML-FeSi in enabling high temperature operation in spintronics.

Generally, materials with low in-plane stiffness are preferred to enhance their suitability for flexible applications. The in-plane stiffness can be determined using the 2D Young's modulus with the following equation

$$Y_{2D} = A_0 \left(\frac{\partial^2 E_s}{\partial A^2}\right)_{A_0}, \qquad (3)$$

where $A$ and $E_s$ are the surface area and the corresponding total energy, respectively. Figure 4(b) shows the profile of $E_s$ versus $A$ for ML-FeSi, which deviates slightly from the typical quadratic relation due to its unique response to isotropic tensile and compressive strains, but can be well described by a cubic polynomial. By taking the second derivative of the curve at $A_0$, we obtain $Y_{2D}$=14 N/m. This value, to our knowledge, is significantly lower than that of any other flat 2D crystal reported so far. Therefore, this is the first time that a combination of large perpendicular magnetic anisotropy, high FM $T_c$, ferroelasticity, and flexibility has been observed in a crystalline material at the 2D limit.

As mentioned before, the synthesis of ML-FeSe has triggered rich properties in this family of materials. It is natural to wonder whether the demonstration of ML-FeSi could lead to the discovery of a material family with appealing multifunctionalities, or whether some of them can exhibit FEL and/or good flexibility. In addition, are there any commonalities among these individuals? To address these questions, an extensive investigation of several isostructural binary transition metal compounds was conducted. These crystals include FM MnGe, MnP, MnAs, CrP, CrAs, FeI, VP, VAs, TiTe, AFM FeSe, and nonmagnetic α-SnO and α-PbO MLs, 12 in total.

The lattice constant, ground state, MAE, and easy axis of this material family were investigated. For MnGe, MnP, MnAs, CrP, CrAs, FeI, and FeSe MLs, the basic properties are verified to align with previous studies conducted at the same computational level [35,36,38]. But crucially, contradicts previous calculations [30,37], the in-plane magnetization is found to be more stable than the corresponding out-of-plane magnetization in energy by 78, 123, and 709 μeV per TM atom for VP, VAs, and TiTe, respectively, implying an easy-plane isotropic magnetic polarization in their FM ground states [Fig. 5(a)]. Therefore, they should be classified as XY magnets, with the absence of an FM phase transition. In contrast, they will transition to a quasi-long-range ordered low-temperature phase through a BKT transition. Using the classical XY model with nearest-neighbor coupling, the critical temperature for the BKT transition is given implicitly by $T_{BKT}$=0.89$J_1/k_B$, where $k_B$ is the Boltzmann constant [10,11,64]. Our calculations yielded $T_{BKT}$ values ranging from 260 K for ML-MnGe to 665 K for ML-TiTe, as summarized in Table I. The additional spin degree of freedom combined with the high $T_{BKT}$ value makes these systems ideal for studying nontrivial topological spin states. This result underscores the influence of composition on the distinct zero-temperature ground-state magnetization.

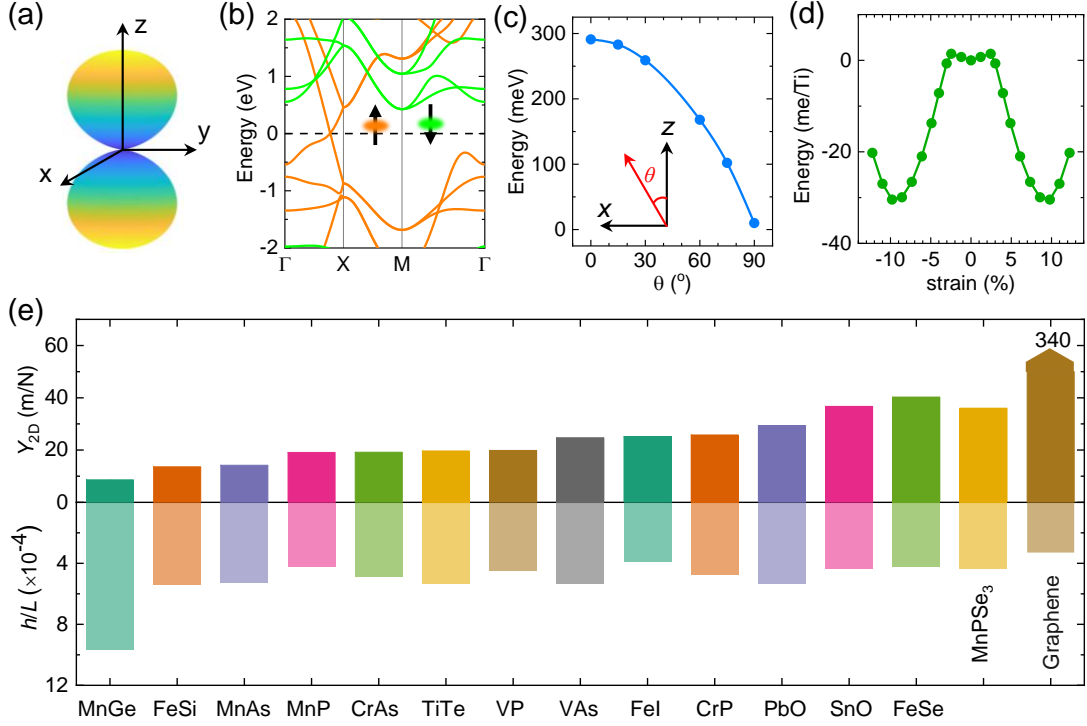

FIG. 5. (a) MCA in VP, VAs, and TiTe MLs. (b) Band structure of ML-TiTe in the absence of SOC by using HSE06 method. The symbols "↑" and "↓" denote up and down spins, respectively. (c) Dependence of the SOC gap of ML-TiTe on the spin orientation quantified by a polar angle $\theta$. (d) Elastic energy per atom of ML-TiTe as a function of in-plane antisymmetric diagonal strain. (e) In-plane stiffness and gravity-induced deformation $h/L$ of various 2D crystals resembled to ML-FeSi in comparison with values reported for ML-MnPSe$_3$ and graphene.

Furthermore, the electronic band structure of ML-TiTe without considering the SOC interaction shown in Fig. 5(b) gives an ideal, fully spin-polarized 2D Dirac semimetal state, which is the parent state of a QAH insulator [32]. When the SOC is included, the Dirac point is gapped (~292 meV). By altering the out-of-plane magnetization towards an in-plane direction, the Dirac gap diminishes progressively until it hits zero [Fig. 5(c)]. This indicates that the intrinsic easy-plane polarization serves as a pivotal point in switching the QAH state, along with a change in the sign of the Chern number and a reversal in the direction of propagation of the chiral edge state. Also, it is found that ML-TiTe also exhibits a FEL behavior [Fig. 5(d)], very similar to that of FeSi, α-SnO, and α-SnO MLs. As a common feature of easy-plane ferromagnets, the bounded in-plane magnetic anisotropy and the lattice direction in monolayer TiTe allows the manipulation of MAE through a strain-induced FEL transition [23-26].

In addition to ML-FeSi, our numerical results demonstrate that all of these systems are soft (Figs. 5(c) and S8 in the Supplemental Material [34]), with the corresponding small $Y_{2D}$ values listed in Table I. For comparison, the $Y_{2D}$ of graphene was calculated using the same method, yielding 334 N/m, in perfect agreement with experiment (340 N/m) [65,66] and theoretical prediction (335 N/m) [67]. These results indicate the reliability of our calculations and highlight the important role of the lattice type in determining mechanical preference. The novel properties (including magnetism and ferroelasticity) in this material family are challenging to replicate in alternative 2D systems characterized by conventional hexagonal and honeycomb lattices.

TABLE I. The lattice constant ($a=b$, in Å), magnetic anisotropy energy (in μeV/TM), easy-axis (EA), estimated critical transition temperature ($T_c$ or $T_{BKT}$, in K), Young's modulus $Y_{2D}$ (in N/m), and deformation $h/L$ for several 2D TMX crystals. All calculations (except FeSe, TiTe, α-SnO, and α-PbO MLs) were performed using the PBE+$U$ method, with the $U_{eff}$ values set to 3 eV for TM ions. According to previous reports, the calculations of the TiTe, FeSe, α-SnO, and α-PbO MLs were performed at the PBE level.

| Systems | $a$ | MAE | EA | $T_c$, $T_{BKT}$ | FEL | $Y_{2D}$ | $h/L$ (×10$^{-4}$) |
|---------|-----|-----|-----|------|-----|------|------|
| FeSi | 3.60 | 580 | z | 830 | yes | 14.0 | 5.38 |
| MnGe | 4.31 | 580 | xy | 260 | yes | 8.60 | 9.71 |
| MnAs | 4.54 | 244 | z | 1000 | no | 14.17 | 5.25 |
| MnP | 4.43 | 128 | z | 790 | no | 19.07 | 4.21 |
| CrAs | 4.29 | -230 | xy | 438 | no | 19.16 | 4.89 |
| CrP | 4.20 | 27 | z | 1400 | no | 25.17 | 3.93 |
| VP | 4.47 | -78 | xy | 263 | no | 19.85 | 4.47 |
| VAs | 4.62 | -123 | xy | 438 | no | 25.77 | 4.43 |
| FeI | 3.89 | 600 | z | 530 | no | 24.73 | 5.35 |
| TiTe | 4.08 | -709 | xy | 665 | yes | 19.64 | 5.37 |
| FeSe | 3.68 | 0 | - | - | no | 40.25 | 4.24 |
| α-SnO | 3.84 | 0 | - | - | yes | 36.70 | 4.33 |
| α-PbO | 4.03 | 0 | - | - | yes | 29.33 | 5.34 |

As reported, closely related materials FeSe ML and bilayer have been broadly produced using the MBE method or traditional mechanical and Al$_2$O$_3$-assisted exfoliation techniques [27,47,68,69]. Due to the moderate lattice mismatch between the TMX MLs (3.60~4.62 Å) and STO (3.91 Å), there is a significant opportunity to fabricate them in experimental settings. On the other hand, according to previous reports and our confirmation, most of the bulk ATM$_2$X$_2$ have similar $E_{cl}$ values to AFe$_2$Si$_2$, indicating the feasibility of obtaining other TMX MLs (except for FeI and TiTe as no bulk phase exists for them to date) from their bulk counterparts through mechanical exfoliation.

To form a stable 2D structure experimentally, a ML must be able to hold its own weight and maintain its geometry [70]. Elastic theory indicates that the usual out-of-plane deformation, $h$, resulting from gravity can be determined using the formula [71]

$$\frac{h}{L} \approx \left(\frac{\rho g L}{Y_{2D}}\right)^{1/3} \quad (4)$$

where $\rho$ is the density of the nanosheets and $g$ is the gravitational constant. For convenience, a substantial flake with a length of $L$=100 μm is assumed here for comparison to previous researches. Remarkably, as shown in Fig. 5(e), the estimated $h/L$ values for all the TMX MLs studied are on the order of $10^{-4}$, with detailed values also listed in Table I. These values are comparable to typical 2D materials such as ML-MnPSe$_3$ ($4.4\times10^{-4}$) [52] and graphene ($\sim3\times10^{-4}$) [70]. Thus, they could have sufficient strength and structural integrity to prevent curling and form a freestanding, self-supporting large area membrane. The theoretical conclusions drawn in this research may encourage additional experimental studies on this distinct class of materials.

Based on first-principles calculations and MC simulations, we have proposed an emergent 2D exfoliated FeSi nanosheet with a square lattice. This 2D crystal possesses an FM-FEL multiferroicity with an estimated $T_c$ of up to 830 K, much higher than the highest experimental value achieved for ML-Cr$_3$Te$_4$. Distinct from the conventional hexagonal and honeycomb lattices, our results reveal that ML-FeSi is unexpectedly flexible with a Young's modulus much lower than that seen in previous works on 2D atomic crystals with flat surfaces. Importantly, the 2D lattices in MnP, MnAs, CrP, CrAs, MnGe, FeI, VP, VAs, and TiTe are characterized by being soft intrinsic ferromagnets with a preferred polarization, either out-of-plane or easy-plane. These novel features are likely facilitated by the combined effects of the crystal structure and the presence of electron-deficient transition-metal ions. We also demonstrate FM-FEL multiferroicity in ML-TiTe, where both the magnetic polarization and the topological features can be switched by elastic strain. The simultaneous emergence of high-temperature ferromagnetism, flexibility, and/or ferroelasticity within a single material is intriguing and is expected to provoke further experimental and theoretical investigations. Our work endows this unique material family with great potential in functional devices, and provides the ideal platform for studying nontrivial topological spintronics.